\documentclass[aps,prl,twocolumn,superscriptaddress]{revtex4-1}

\usepackage{graphicx}
\usepackage{xcolor}

\begin{document}

\title{Limits on the flux of nuclearites and other heavy compact objects from the ``Pi~of~the~Sky'' project}

\author{Lech Wiktor Piotrowski}
\email[]{publ@lwp.email}
\affiliation{RIKEN, Wako, Japan}

\author{Katarzyna Ma{\l}ek}
\affiliation{National Centre for Nuclear Research, ul. Pasteura 7, 02-093 Warsaw, Poland}
\affiliation{Aix Marseille Univ. CNRS, CNES, LAM, Marseille, France}

\author{Lech Mankiewicz}
\affiliation{Centre for Theoretical Physics, Polish Academy of Science, Warsaw, Poland}

\author{Marcin Soko{\l}owski}
\affiliation{International Centre for Radio Astronomy Research, Curtin University, Bentley, WA 6102, Australia}

\author{Grzegorz Wrochna}
\affiliation{National Centre for Nuclear Research, ul. Pasteura 7, 02-093 Warsaw, Poland}

\author{Adam Zadro\.zny}
\affiliation{National Centre for Nuclear Research, ul. Pasteura 7, 02-093 Warsaw, Poland}

\author{Aleksander Filip \.Zarnecki}
\affiliation{University of Warsaw, Warsaw, Poland}

\date{\today}

\begin{abstract}
	Many theories predict the existence of very heavy compact objects, that in terms of sizes would belong to the realms of nuclear or atomic physics, but in terms of masses could extend to the macroscopic world, reaching kilograms, tonnes or more. If they exist, it is likely that they reach our planet with high speeds and cross the atmosphere. Due to their high mass to size ratio and huge energy, in many cases, they would leave behind a trail in the form of sound and seismic waves, etches, or light in transparent media. Here we show results of a search for such objects in visual photographs of the sky taken by the ``Pi of the Sky'' experiment, illustrated with the most stringent limits on the isotropic flux of incoming so-called nuclearites, spanning between $5.4\cdot10^{-20}$ and $2.2\cdot10^{-21}\ \mathrm{cm}^{-2} \mathrm{s}^{-1} \mathrm{sr}^{-1}$ for masses between 100 g and 100 kg. In addition we establish a directional flux limit under an assumption of static ``sea'' of nuclearites in the Galaxy, which spans between $1.5\cdot10^{-18}$ and $2.1\cdot10^{-19}\ \mathrm{cm}^{-2} \mathrm{s}^{-1}$ in the same mass range. The general nature of the limits presented should allow one to constrain many specific models predicting the existence of heavy compact objects and both particle physics and astrophysical processes leading to their creation, and their sources.
\end{abstract}

\maketitle

\section{Introduction}

The most extreme case of a heavy compact object that has ever been detected experimentally is a black hole -- an object so heavy that it packs all its mass in a possibly infinitely small amount of space. The ones that have been observed so far are extremely heavy. Not much prevents, however, the existence of much lighter counterparts. They are just but one example of many types of heavy compact objects predicted by different branches of physics and astrophysics.

One other example is nuclearites -- a name usually attached to heavy strangelets, hypothetical lumps of ``strange quark matter'' predicted by Witten\cite{bib:witten}, consisting of roughly equal numbers of up, down and strange quarks and being more stable than the ordinary matter consisting of only up and down quarks. De Rujula and Glashow predicted\cite{bib:nucl_atm} that nuclearites, travelling with speeds of the order of 100 km/s would collide elastically with atoms. Their energy loss mechanism would be similar to that of a meteor, however, their compact size would allow those heavier than $4\times 10^{-14}$ g to penetrate the atmosphere, and those heavier than 0.1 g to pass through the Earth's diameter. In the process of traversing through a transparent medium such as air or water, they would create an expanding thermal shock-wave and thus convert part of their energy into visible light. The amount depends mainly on the density of the medium, and speed and radius of the object.

The reasoning developed for nuclearites can be applied to different hypothetical compact objects that could interact with atoms in a similar manner. The most notable candidates are Q-Balls\cite{bib:q-balls} and magnetic monopoles\cite{bib:monopoles}. However, the list of possibilities is much longer, including fermionic exotic compact stars\cite{bib:fecs}, primordial black holes\cite{bib:pbh} and their remnants\cite{bib:pbh-remnants}, mirror matter\cite{bib:mirror-matter}, Fermi balls\cite{bib:fermi-balls}, electroweak symmetric dark matter balls\cite{bib:esdmb}, anti-quark nuggets\cite{bib:aqn}, axion quark nuggets\cite{bib:axqn}, six-flavour quark matter\cite{bib:6fqm} and non-strange quark matter\cite{bib:nsqm}. The details of interactions with ordinary matter have to be studied separately for each hypothesis.

For many of these candidates, including nuclearites, the light emitted in the atmosphere would create a light trail similar to that of meteors, but mostly in the lower atmosphere and reaching the ground. They would have a small loss of energy, no loss of mass and nearly constant speed, and thus produce a track with almost constant absolute brightness. In addition, Solar System meteors do not exceed speeds of 72 km/s, while in the scenarios of massive compact objects being of galactic or extragalactic origin, bound to the Galaxy as Dark Matter (DM), or coming from collisions or explosions of astrophysical objects, their speeds in most cases would be at least a few times higher. Despite the differences, they would be observable by star-gazing with on-ground telescopes. Those monitoring large parts of the sky, like the ``Pi of the Sky'' experiment, would be the most likely to detect them.

\section{The ``Pi of the Sky'' experiment}

The ``Pi of the Sky'' experiment\cite{bib:burd} was a~system of wide field of view robotic telescopes designed to search for variable astrophysical phenomena. The design of the apparatus allowed to monitor a~large fraction of the sky with a range of $12^{m} - 13^{m}$\footnote{Magnitude $^{\rm{m}}$ stands for an apparent brightness of an astronomical object in relation to an object of reference: $m=-2.5\log{\frac{I}{I_r}}+m_r$, where $m_r$ is the magnitude of the object of reference, $I_r$ its observed radiation flux and $I$ is the observed radiation flux of the discussed object. This unit has been normalized according to the Ptolemy scale, where the brightest stars seen with the naked eye had a brightness of $1^{\rm{m}}$ and the darkest of $6^{\rm{m}}$. The Sun has a brightness of $-26^{\rm{m}}$, and Vega $0^{\rm{m}}$.} (these values are not used for obtaining the results presented in this manuscript, as they are too general) and time resolution of the order of $10$~seconds.  The main goal of the project was to search for optical counterparts of Gamma Ray Bursts (GRBs) during or even before gamma-ray emission\cite{bib:grb_naked,bib:grb_160625b}. 

The experiment was equipped with custom-designed cameras and Canon lenses with focal length f = 85~mm, f/d = 1.2 (d standing for the diameter of the entrance pupil), each camera covering $\sim 20^{\circ}\times 20^{\circ}$ of the sky with roughly 4 million pixels. The full system consisted of 16 cameras placed on equatorial mounts (4 cameras per mount), covering almost 2 sr of the sky, working since 2013 in the INTA El Arenosillo test centre in Mazag\'on near Huelva, Spain. Before that, a~prototype consisting of 2 cameras working in coincidence and observing the same field of view had been working at Las Campanas Observatory in Chile since June 2004 until the end of 2009, and later in San Pedro de Atacama in Chile. The project stopped gathering data in 2017.

\section{Search for compact heavy objects}

The 10 s exposure time of the ``Pi of the Sky'' cameras is not optimised for fast-moving heavy compact objects. The time spent in any of the camera's pixels is very short compared to the whole exposure and decreases the signal to noise ratio, and thus limits the sensitivity to low mass, dim events. However, the experiment was monitoring a very large part of the atmosphere and gathered a significant amount of data, useful for looking for small fluxes.

Using a custom track search algorithm (see \cite{bib:methods}), we have analysed our archived raw data: 337674 frames with 10 s exposure and 34004 stacked frames, consisting of $20 \times 10$ s exposures from many locations on the sky. Among those 185258 single and 22237 stacked frames were left after quality cuts and a requirement that the centre of the field of view is pointing more than $20^{\circ}$ above the horizon. This adds up to 1766.05 h of clear observation for a single camera equivalent.

After performing track detection and initial removal of spurious events (clouds, cosmic rays, etc.) on the frames, 35870 tracks (corresponding to the 1766.05 h residual observation
 time) passed on to the next analysis stage. Among those, 33257 were automatically classified as meteors or satellites and 2613 underwent manual inspection. 

The manual inspection consisted of analysing intensity profiles and track images. The main challenge is to distinguish between nuclearites and meteors/satellites. The change in intensity of a nuclearite signal comes only from the changing distance to the detector and is thus expected to be small, very smooth and slow. Therefore all events with rapid changes in brightness have been classified as meteors or satellites. The inspection left 29 candidates for nuclearites.

The 29 remaining tracks were compared with the NORAD satellite database for the corresponding nights and 9 events were identified as satellites, which left us with 20 candidates for nuclearites. However, the NORAD satellite database is far from complete, and it is quite likely that many other of those events are satellites.

\begin{figure*}
	\begin{center}
		\includegraphics[width=0.49\textwidth]{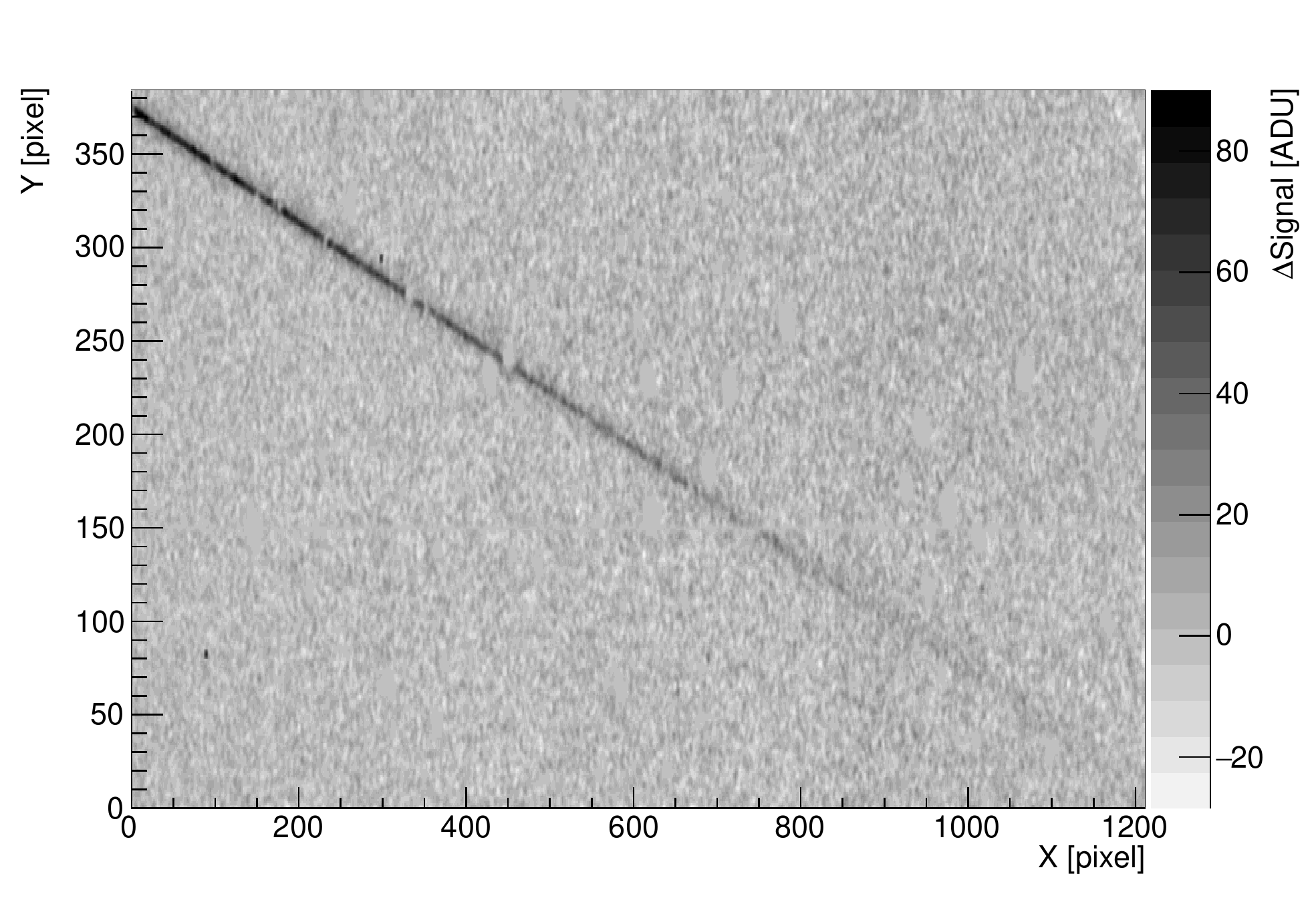}
		\includegraphics[width=0.49\textwidth]{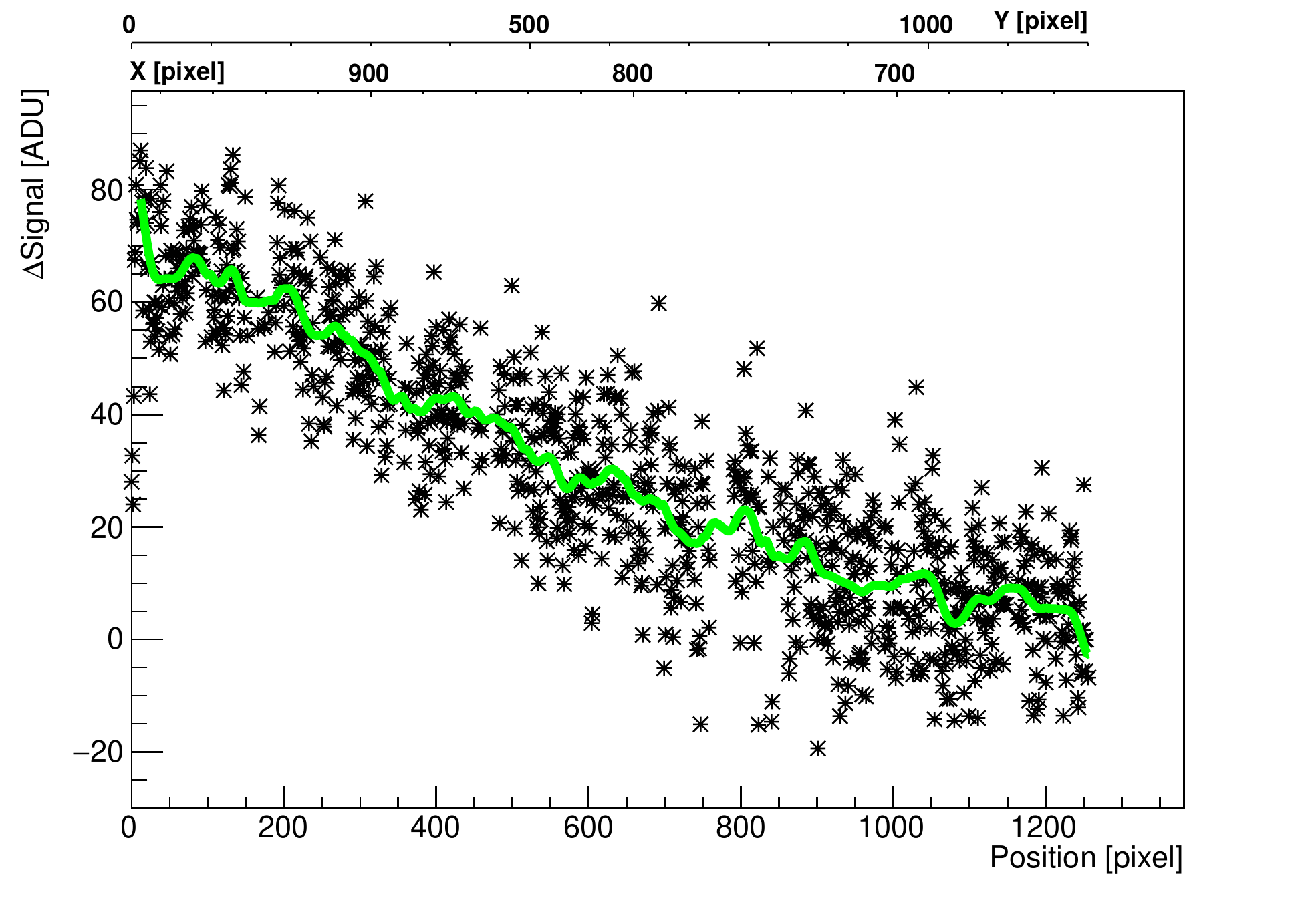}
		\caption{The longest of the remaining nuclearite candidates. Left: the track on the frame after subtraction of the previous frame and Gaussian smoothing. Right: the intensity profile of the event, with the curve showing the Gaussian smoothing of the measurement points.}
		\label{fig:candidate}
	\end{center}
\end{figure*}

None of the 20 events is an obvious candidate for a nuclearite track, which should be very long, with an almost constant and preferably strong signal. The longest one is about 1055 pixels, exiting the frame (fig. \ref{fig:candidate}). One could speculate about multiple trends in the presented intensity profile which should not exist for a real nuclearite, however definite exclusion of this event is hard.

\section{Limits on the flux of heavy compact objects}

To estimate the limits on the flux of heavy compact objects in the ``Pi of the Sky'' experiment, we need to establish the maximal achievable flux for the selection of object masses and multiply it by the efficiency of detection and separation from other types of tracks in the atmosphere. For this purpose we have simulated nuclearites crossing our field of view with brightness in stellar magnitude units:

$$\mathcal{M}=10.8-1.67\log_{10}(m/1\ \mu g)+5\log_{10}(d/10\ km)$$

where $m$ is the mass and $d$ is the distance to the telescope. The maximal altitude at which nuclearites effectively generate light is:

$$h_{max}=2.7\ km\ \ln(m/1.2\times 10^{-5}\ g)$$

both equations following the calculations of De Rujula and Glashow\cite{bib:nucl_atm}. Next, we applied our detectors' parameters such as exposure, PSF, pointing zenith angle, atmospheric extinction, etc. Using this procedure we have determined the detector's effective surface (including detection efficiency) and thus calculated limits for an isotropic flux and directional flux caused by Earth's movement in a ``sea'' of a static halo of nuclearites.

\subsection{Isotropic limiting flux}

The isotropic flux of heavy compact objects could come from extragalactic sources such as GRBs or Galactic sources with isotropic distribution around the Earth. It is also used as an approximation of a flux of Dark Matter objects by De Rujula and Glashow following the isothermal sphere assumption of the Dark Matter's Standard Halo Model (SHM).

\begin{figure*}
	\begin{center}
		\includegraphics[width=0.6\textwidth]{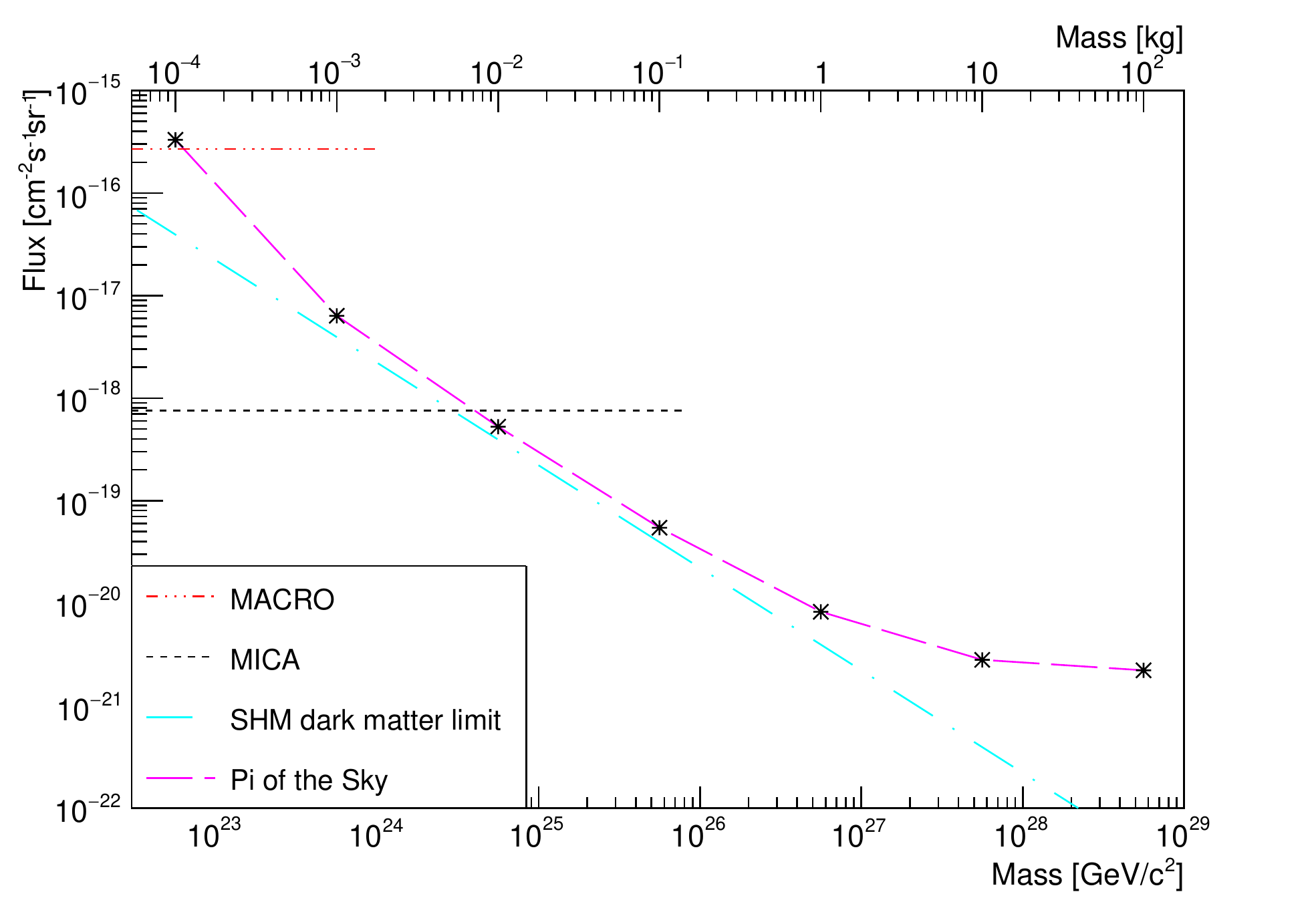}
		\caption{90\% CL limit for the isotropic flux of nuclearites of specific mass by the ``Pi of the Sky'' project on top of the limits given by MACRO and a mica minerals analysis. We show also the constraint in the SHM dark matter scenario, assuming its isotropic flux on Earth.}

		\label{fig:flux_limit}
	\end{center}
\end{figure*}

We have simulated a random isotropic flux of nuclearites crossing the volume of the atmosphere observed and then supplied the results to the detection algorithm, to estimate our detector's effective area $S_E^{Iso}(m)$, which is approximated by the surface of the field of view pyramid of a camera and depends on nuclearite mass $m$, sky conditions, and the detector's pointing zenith angle. This results in the following formula for the 90\% CL (confidence level) limit on the isotropic flux:

\begin{equation}
\label{eq:iso_flux}
\Phi(m) = \frac{2.3}{S_E^{Iso}(m)\cdot t_e \cdot 2\pi}
\end{equation}

where $t_e$ stands for the exposure time (10 s for single and 200 s for stacked frames) and $2\pi$ comes from the fact that we do not take into account nuclearites coming from below the detector. Fig. \ref{fig:flux_limit} shows the obtained flux limit on top of the limits given by MICA \cite{bib:mica} and MACRO\cite{bib:macro}. It is important to mention that SLIM has given a limit of $\sim 10^{-15}$ cm$^{-2}$s$^{-1}$sr$^{-1}$ for nuclearites with mass below $10^{21}$ GeV/c$^2$ \cite{bib:slim}, and ANTARES a limit of $\sim 10^{-17}$ cm$^{-2}$s$^{-1}$sr$^{-1}$ for nuclearites with mass below $10^{17}$ GeV/c$^2$ \cite{bib:antares}.

\subsection{Directional limiting flux}

\begin{figure*}
	\begin{center}
		\includegraphics[width=0.6\textwidth]{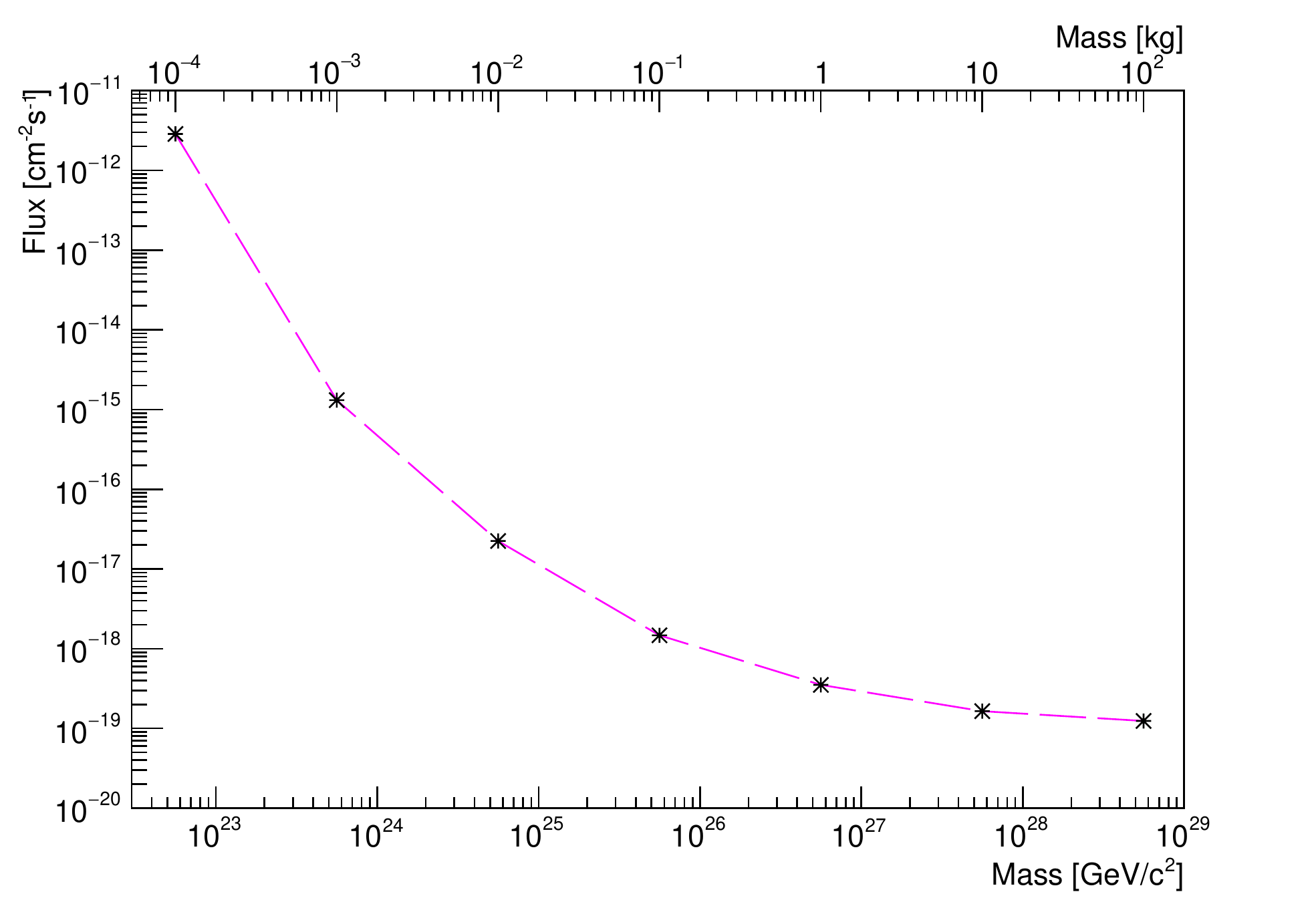}
		\caption{90\% CL limit for the directional flux of nuclearites of specific mass by the ``Pi of the Sky'' project}
		
		\label{fig:flux_directional_limit}
	\end{center}
\end{figure*}

Here we consider a case where nuclearites are static in the Galaxy and bombard the Earth due to its movement through the Milky Way along with the Solar System. This can be used as a basis for calculations involving more complicated assumptions about distributions and velocities of compact heavy objects. In this case, the flux will not be isotropic but aligned along the telescope's velocity vector. The 90\% CL limit on the directional flux is given by:

\begin{equation}
\label{eq:dir_flux}
\Phi(m) = \frac{2.3}{S_E(m)\cdot t_e}
\end{equation}

losing the $2\pi$ factor from the isotropic flux. The effective surface $S_E(m)$ includes the field of view pyramid's cross-section perpendicular to the flux direction for a specific frame. The flux limit is drawn in fig. \ref{fig:flux_directional_limit}. Only 142772 single and 7984 stacked frames where the flux was coming from above the horizon were taken into account.

\section{Discussion of the results}

Presented results can be applied to heavy compact objects, assuming they enter the atmosphere with a speed of the order of hundreds of kilometres per second and interact with the atoms elastically or semi-elastically. The assumption is based on the well-known behaviour of meteors and orbital objects during orbital re-entry. The mass-scale, selected for nuclearites, assumes that the cross-section is purely geometrical. The results can be used for different objects after the mass-scale is adjusted taking into account their cross-section or different light emission mechanisms.

The flux limit value is determined mainly by the experiment's field of view, exposure time and number of frames analysed. The estimation of the detector's efficiency in detecting objects of specific mass plays a secondary role in the presented mass-region, and possible deviations from estimated values due to simulation uncertainties would hardly be visible on the logarithmic flux scale. Only downward going objects were taken into account. The limits could be simply divided by 2 for 1 kg and heavier nuclearites, if upward going objects were to be considered, as in this mass range traversing the Earth causes almost no speed loss. The selection efficiency could be improved with a detailed analysis of the brightness of the selected events and comparing them to specific objects masses.

The presented mass range could be extended to higher masses until tracks become so bright as to saturate the detector. Assessing when this happens is out of the scope of this paper, but it is clear from the presented curve, that the limit would not change significantly (strengthening slightly due to the increasing maximal altitude at which nuclearites effectively generate light). With enough time and computing power invested into further simulations, we could extend the limits to lower masses, but due to extremely small efficiencies, the possibility of detection would rely on a very small chance that a heavy compact object passes very close to the detector.

\subsection{Constraints on Dark Matter}

De Rujula and Glashow estimated an upper bound on the flux of nuclearites as an isotropic flux of Dark Matter on Earth given by $F=7.8\cdot (1\ \mathrm{g}/m)$ (where m is the mass of a nuclearite) based on the local DM density of the SHM. Our isotropic limits are weaker than this bound, with the closest point being $\sim 60\%$ above (unless uncertainties of the SHM are considered). Fine-tuning our analysis, such as taking into account upward going nuclearites, would put us on the other side of the upper bound line. However, even in this case, we would not attempt to put constraints on the SHM. One has to realise that the upper bound assumes a flux of the same mass objects. This could be true for an elementary particle, but not for a bag of quarks with an unknown, but probably mostly continuous spectrum. Also, it is important to mention that the local dark matter density
 is uncertain within a factor of a few, and the SHM approximation
 may prove overly simplistic. Finally, the assumption of an isotropic flux of DM composites on the Earth requires the planet to be stationary in the Galaxy. Therefore we find our limits, especially the directional limit as a good basis for limit calculations for specific compact objects and their expected flux distributions, and the isotropic limits for non-DM origin scenarios.

It is worth noting that for objects with a higher interaction cross-section, the limit curve shifts towards lower masses. For example, for magnetised nuclearites\cite{bib:magnetised_nuclearites}, the shift is about 9 orders of magnitudes, allowing for constraining the SHM.

In addition, the limits on the flux of heavy compact objects can be transformed into limits on the cross-section under several assumptions related to the SHM, as shown by Sidhu and Starkman \cite{bib:sidhuandstarkman} in their limit estimates based on an interpretation of bolide camera networks observations. Cross-section limits coming from the Pi of the Sky data will be a subject of a separate publication.

\section{Conclusions}

We have shown flux limits on compact heavy objects, in the mass scale range considered for nuclearites. These are the most stringent limits up to date in the 100 g -- 100 kg range according to the authors' knowledge. The limits can be extended to much higher masses until the object becomes too bright to be recognised as a track. The isotropic limit can be used mainly under the assumption of extragalactic sources such as short GRBs, which, in case they are colliding strange stars could eject nuclearites, which in turn, if beamed like electromagnetic radiation, could perhaps supply a significant flux on the Earth. The directional flux can serve as a basis for calculations involving specific scenarios of spatial and velocity distributions of objects and could be useful in investigations of dark matter hypotheses, especially if the mass scale is shifted towards lower masses, which would be the case for objects with higher cross-section, such as magnetised nuclearites. Finally, these limits may result in constraints on a growing number of astrophysical scenarios, ranging from early Universe evolution to cosmic cataclysms, where the production of heavy compact objects is expected.

\begin{acknowledgments}
	K.M. has been supported by the Polish National Science Centre (UMO-2018/30/E/ST9/00082).
\end{acknowledgments}

\nocite{bib:opencv}
\nocite{bib:unigrid_sphere}
\nocite{bib:scipy}
\nocite{bib:astropy}
\nocite{bib:maximum_gap}
\nocite{bib:signal_length}

\bibliography{paper}
%\bibliography{methods}

\end{document}

\section{Methods}

\subsection{Track search algorithm}

The track search algorithm consisted of the following steps performed for each frame:
\begin{enumerate}
	\item Subtraction of the previous frame from the current, giving a residual image
	\item Detection of line segments on the residual image using the probabilistic Hough transform
	\item Discarding of vertical lines (usually caused by ``bleeding'' -- influence of over-saturated pixels on the pixels in the same column)
	\item Grouping line segments into tracks
	\item Removing spurious tracks caused by clouds
	\item Computing intensity profiles of detected tracks
	\item Storing the parameters and intensity profiles of the remaining tracks for further inspection
\end{enumerate}

The detection of line segments was performed using a probabilistic Hough transform from the OpenCV library\cite{bib:opencv}. While there are more modern and robust methods available, a slight improvement in detection efficiency would not change the limit shown in this paper much, while the computational cost would increase significantly. 

\subsection{Frame selection}

Frames with bad quality such as too bright, too dark, too cloudy or with too big a shift compared to the previous frame were discarded. For this purpose we have used, among others, estimation of the shift using 2D Fourier transform, estimation of the variability separately in 9 parts of the residual image, and custom frame quality estimator -- a standard deviation of the residual image, clipped close to 0, with bright static signal (such as stars) masked. At later steps, frames with an unreasonably high number of line segments or tracks were also discarded. These cuts were made restrictive to leave only clear frames with stable detection efficiency, and discard almost half of the initial data set. In addition, only frames with pointing at least $20^{\circ}$ above the horizon were accepted.

\subsection{False detections}

Nuclearites are not the only track-like events on the frames. Apart from the ``bleeding'' mentioned before, the following are the most numerous ones:

\begin{description}
	\item[Meteors] significantly non-monotonic intensity profile due to burning in the atmosphere. Usually short, for most meteors burn at high altitude.
	\item [Satellites] short (interrupted by start/end of the exposure) intensity profiles with significant flat parts, often visible on several frames
	\item [Airplanes] - accompanied by airplane light flashes
	\item [Cosmic rays] - mostly muons interacting with the CCD sensor, with significant variability of the intensity profile. Often non-linear shape.
\end{description}

\subsection{Nuclearite track simulation}

\begin{figure}
	\begin{center}
		\includegraphics[width=0.46\textwidth]{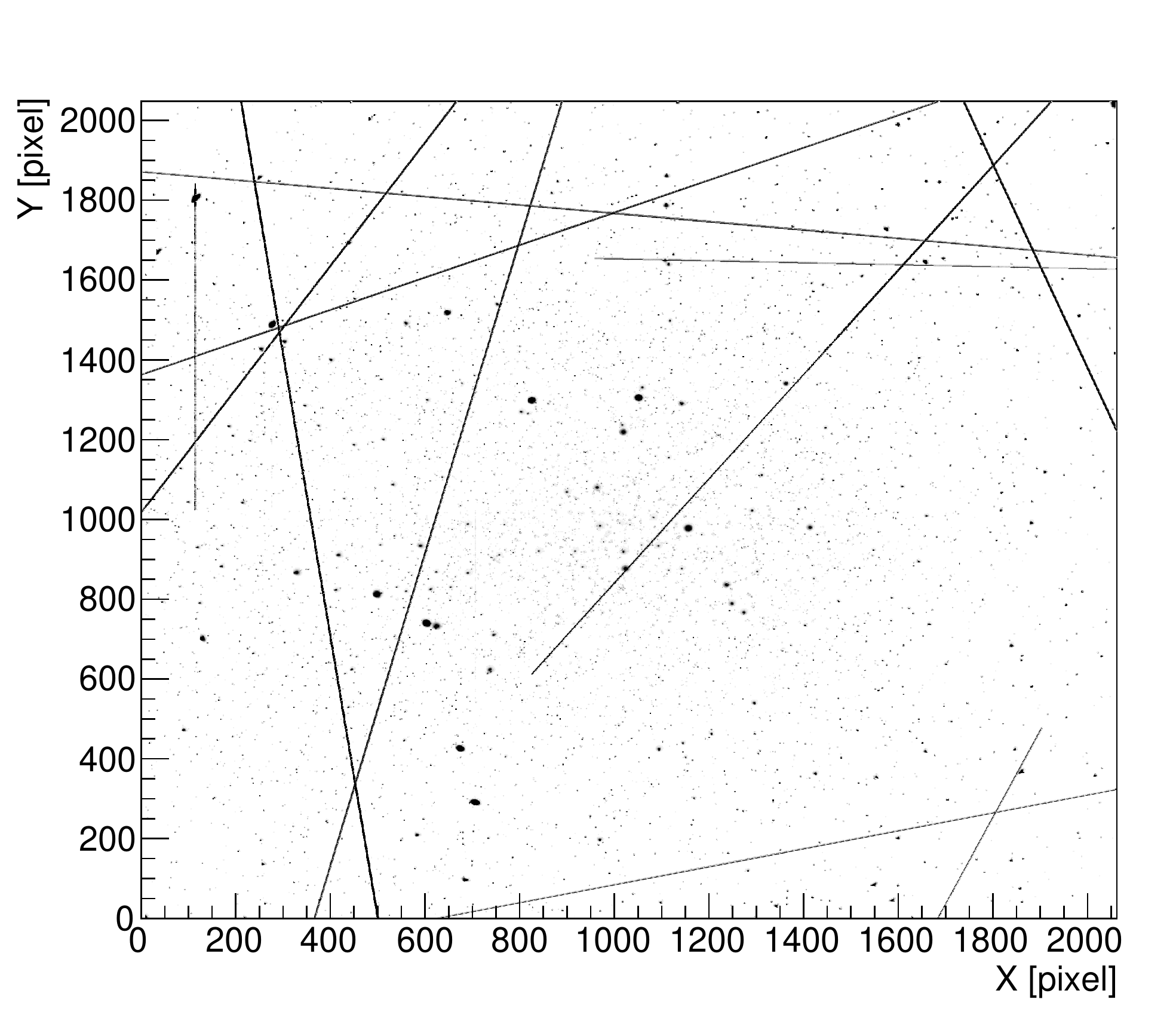}
		\includegraphics[width=0.46\textwidth]{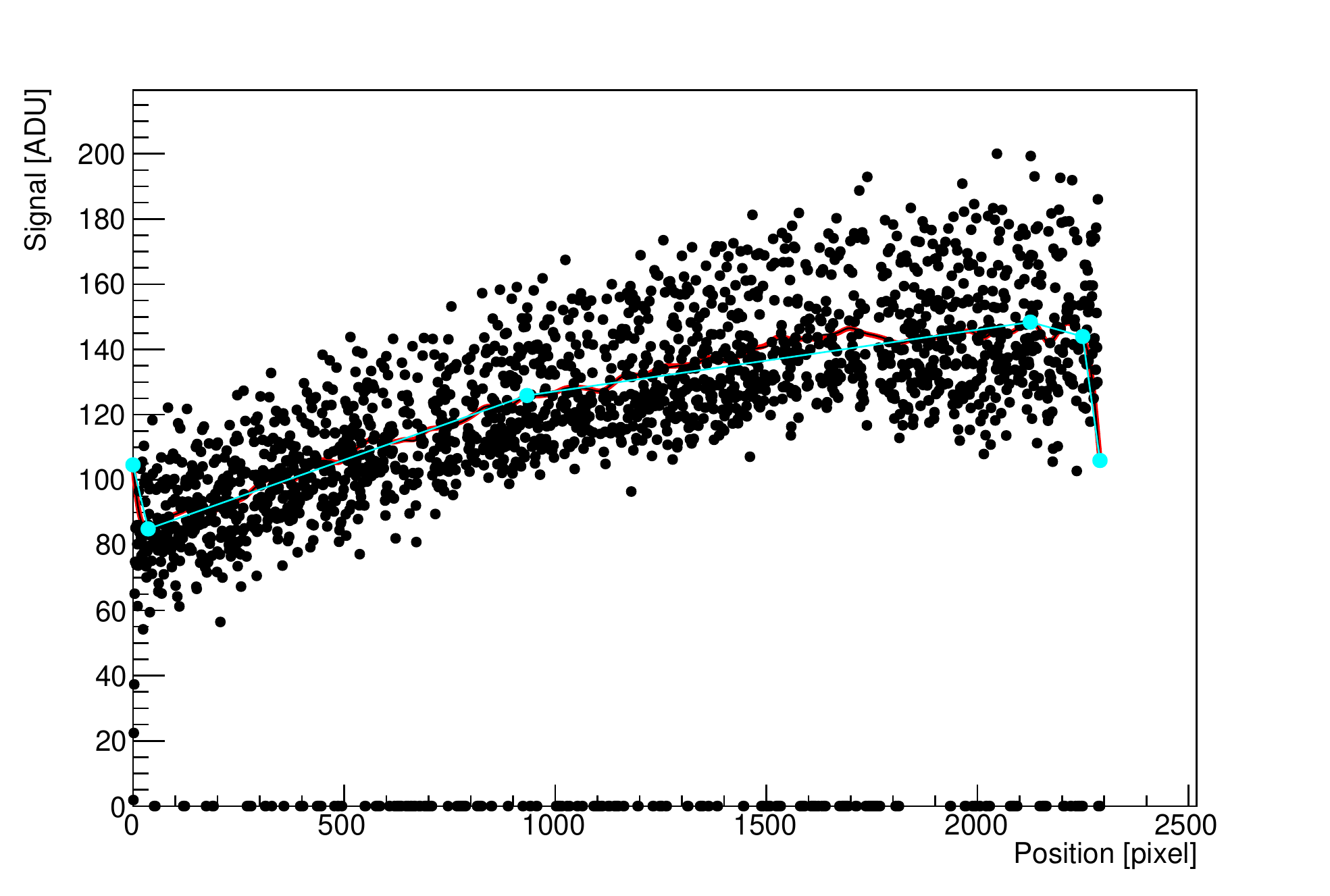}
		\caption{Top: A (brightness thresholded) real sky frame with 10 simulated nuclearite tracks superimposed. Bottom: a simulated nuclearite intensity profile chosen for variability higher than most of the set. The red line shows Gaussian smoothing and the blue one semi-linear trends. Points with zero values belong to the masked parts of the frame (mainly stars positions).}
		\label{fig:simul}
	\end{center}
\end{figure}

Initially, a set of lines in space is generated. The set parameters depend on the simulated flux. For an isotropic flux, random directions are generated. For a directional flux, the set contains tracks with a specific direction in space, at random distances from the detector. Next, only the tracks crossing the pyramid of the field of view (FoV), trimmed by the maximal burning altitude $h_{max}$ for the chosen mass and zenith angle are selected. In this process, the surface of the pyramid in the isotropic case, and the surface of the pyramid's cross-section perpendicular to the flux direction in the directional case are calculated.

The following steps are repeated to estimate the detection efficiency for specific mass, pointing angle and, in the directional case, flux direction:
\begin{enumerate}
	\item Calculate the nuclearite's brightness for the chosen mass:
	\begin{enumerate}
		\item Convert mass to magnitude
		\item Apply atmospheric extinction
		\item Convert magnitude to the camera's pixel counts (ADU)
		\item Reduce the counts according to the nuclearite's time spent in the pixel 
	\end{enumerate}
	\item Convolve the track with the point spread function
	\item Add Poissonian fluctuations to the track
	\item Imprint the track on a real sky frame
	\item Run the track search algorithm
	\item Estimate the detection efficiency
\end{enumerate}

The above procedure is performed repeatedly, using separate sets of randomly selected frames for each pointing angle and camera. The result is an estimation of the detection efficiency for each of the experiment's cameras taking into account changing real background noise, observational conditions, instrument characteristics and effects that we may be unaware of. An example of such a real sky frame with 10 simulated nuclearite tracks superimposed is shown in fig. \ref{fig:simul}.

\begin{figure}
	\begin{center}
		\includegraphics[width=0.46\textwidth]{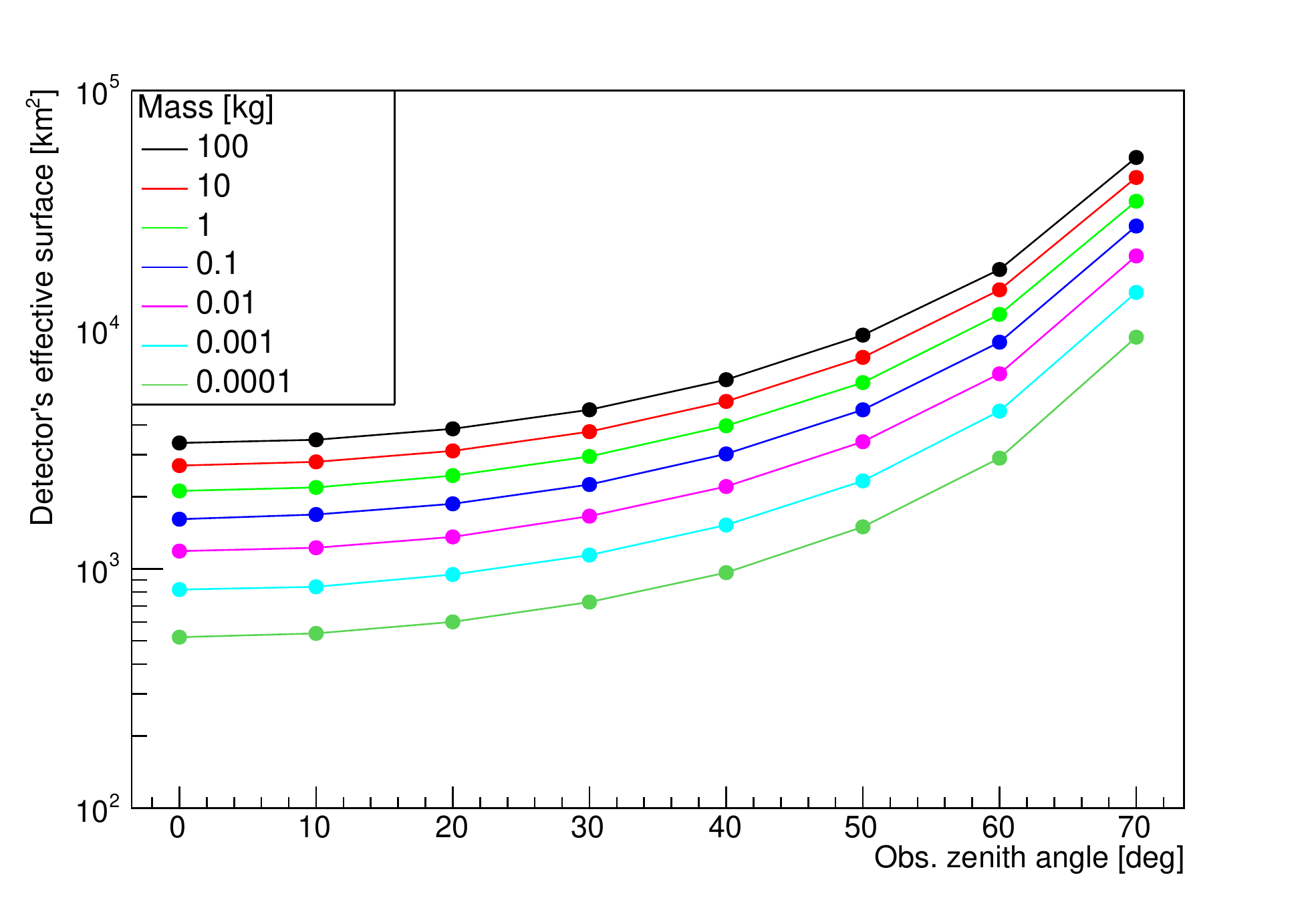}
		\caption{Surface of the FoV pyramid for different masses and the detector's pointing zenith angles, for an observatory at sea level.}
		\label{fig:surf_vs_angle}
	\end{center}
\end{figure}

\begin{figure}
	\begin{center}
		\includegraphics[width=0.46\textwidth]{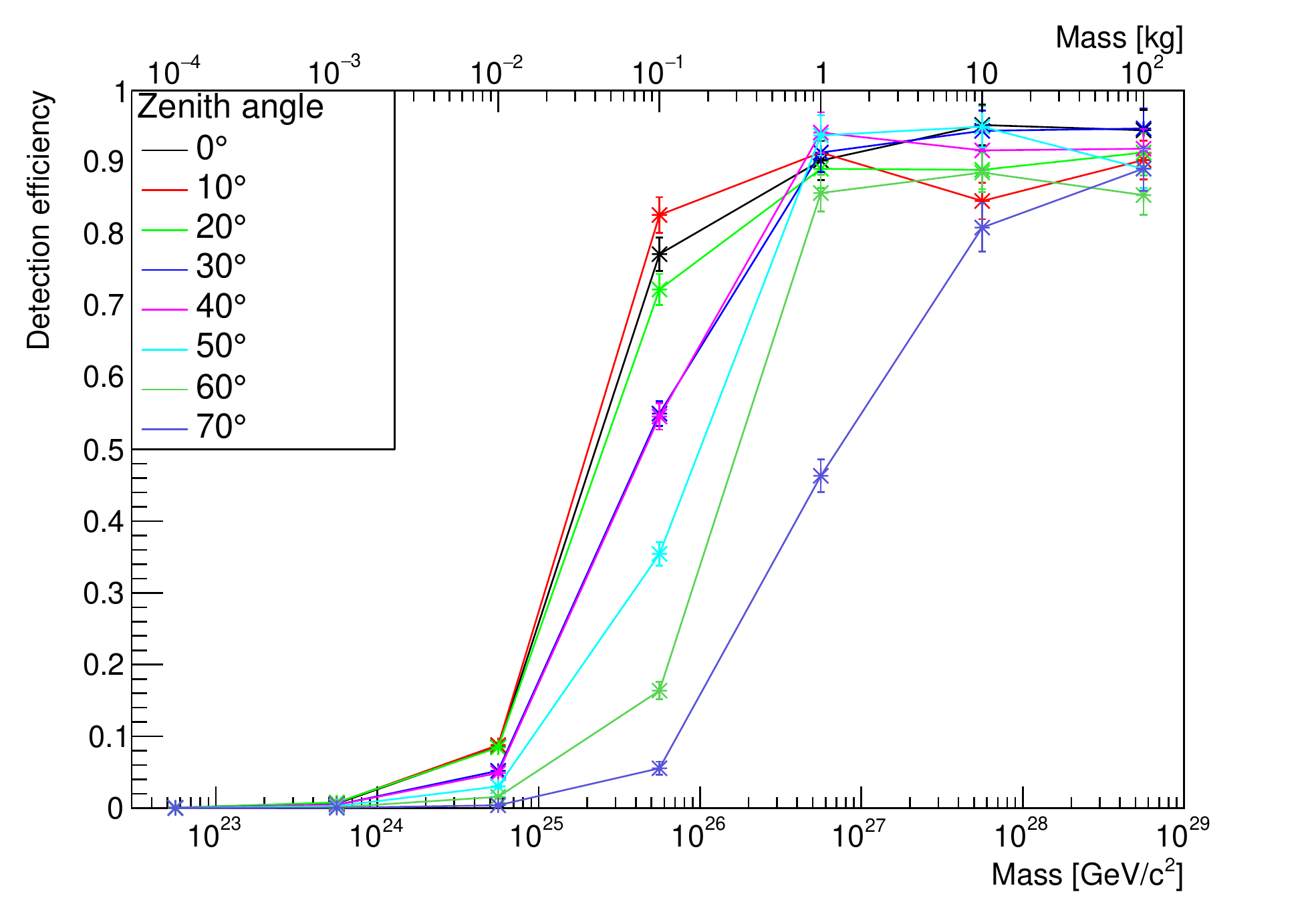}
		\caption{Efficiency of nuclearite detection for different masses and the detector's pointing zenith angles.}
		\label{fig:eff_vs_mass}
	\end{center}
\end{figure}

\subsection{Detector's effective area}

The detector's effective area $S_E(m,\alpha)$ for a chosen mass of the nuclearite $m$ and zenith angle $\alpha$ can be calculated by multiplying the corresponding FoV pyramid surface $s^{FoV}_{m \alpha}$ by the detection efficiency $\epsilon _{m \alpha}$:

$$S_E(m,\alpha)=s^{FoV}_{m \alpha}\cdot \epsilon _{m \alpha}$$

This needs to be multiplied by the number of frames $n_{\alpha}$ taken with specific zenith angle $\alpha$, and summed over all zenith angles to get the final effective area of the detector for the given nuclearite mass used in eq. 1:%\ref{eq:iso_flux}:

$$S_E^{Iso}(m)=\sum_{\alpha =0^{\circ}}^{90^{\circ}} S_E(m,\alpha) \cdot \epsilon _{m \alpha} \cdot n_{\alpha} \cdot \varepsilon_S$$

where $\varepsilon_S$ is an efficiency of separation of nuclearites from other tracks, such as those left by satellites and meteors (see Separation efficiency). 

The FoV pyramid surface $s^{FoV}_{m \alpha}$ has been calculated numerically, and is a fast growing function of the zenith angle $\alpha$, as can be seen in fig. \ref{fig:surf_vs_angle}. It is a direct consequence of the fast growth of the distance to the top of the atmosphere along the optical axis $d$ with $\alpha$:

$$d = \sqrt{(R_E^2\cos^2{\alpha}+2R_Eh+h^2)}-R_E\cos{\alpha}$$

where $h$, the top of the atmosphere at the zenith, is in our case equal to $h_{max}$, the maximum altitude at which a nuclearite of a given mass emits light. $R_E$ is the radius of the Earth. The surface of the pyramid would be proportional to $d^2$ if not for the truncation by the top of the atmosphere $h$.

The detection efficiency $\epsilon _{m \alpha}$ should increase with the track brightness which grows with nuclearite mass, and saturate when almost all tracks are bright enough to be detected. This can be seen in fig. \ref{fig:eff_vs_mass}. On the other hand, the efficiency drops with zenith angle, due to the growing fraction of tracks far from the detector, and thus their reduced intensity. Altogether, the detector's effective area is a result of an interplay between the FoV pyramid surface that grows with the zenith angle and detection efficiency, that drops with the zenith angle. Estimating the optimal zenith pointing angle is out of the scope of this paper, but it was around $30^{\circ}$ and $40^{\circ}$ in our case.

In the directional flux case, the detector effective area for each frame depends on the area of cross section of the FoV pyramid perpendicular to the flux $sc^{FoV}_{m \alpha \phi \theta}$ and the efficiency of nuclearite detection for a given FoV zenith angle $\alpha$, flux direction with respect to the FoV pyramid $(\phi, \theta)$, mass $m$ and separation efficiency $\varepsilon_S$:

$$S_E(m,\alpha,\phi,\theta)=sc^{FoV}_{m \alpha \phi \theta}\cdot \epsilon _{m \alpha \phi \theta} \cdot \varepsilon_S$$

Then the final effective area of the detector for the given nuclearite mass used in eq. 2 is given by:

$$S_E(m)=\sum_{frames}S_E(m,\alpha,\phi,\theta)$$

where we sum over all analysed $frames$, each with defined pointing and the flux direction.

We have performed simulations for six masses (0.1 g, 1 g, 100 g, 1 kg, 10 kg, 100 kg) and ten zenith angles ($0^{\circ}$ to $90^{\circ}$ in $10^{\circ}$ increments). In the directional case this was also combined with 32 $(\phi, \theta)$ pairs forming a uniform grid on the upper hemisphere\cite{bib:unigrid_sphere}. For each specific frame, the above values were linearly interpolated with the SciPy\cite{bib:scipy} interp1d function, and in the case of flux directions, with the SciPy KDTree based algorithm.

The direction of flux for each frame was taken as the vector opposite to the movement of the observatory in the Galactocentric coordinate system for the central time of the exposure and transformed to the observatory's frame of reference using the Astropy package\cite{bib:astropy}. In the process, the speed of the incoming nuclearites was also calculated, and the efficiency modified accordingly (see Speed dependency).

\subsection{Separation efficiency}

\begin{figure}
	\begin{center}
		\includegraphics[width=0.5\textwidth]{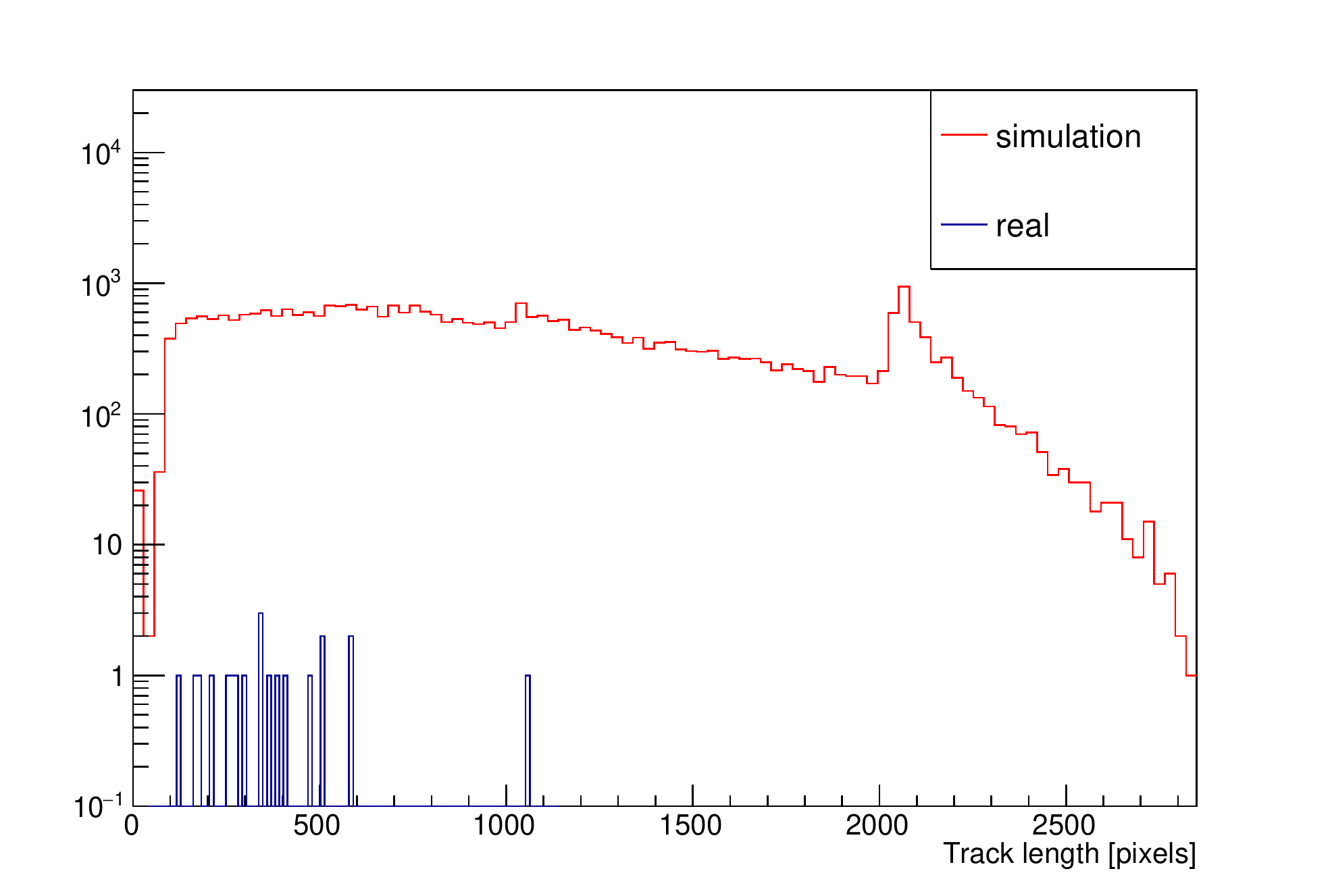}
		\caption{The length of the track of the 20 candidates from the real data and simulated events of all masses. The small peak above 2000 pixels (frame side length) is a geometrical effect caused by an elongation of the field of view pyramid for large zenith angles, and thus slight dominance of tracks parallel to one of the sides.}
		\label{fig:track_length}
	\end{center}
\end{figure}

\begin{figure}
	\begin{center}
		\includegraphics[width=0.5\textwidth]{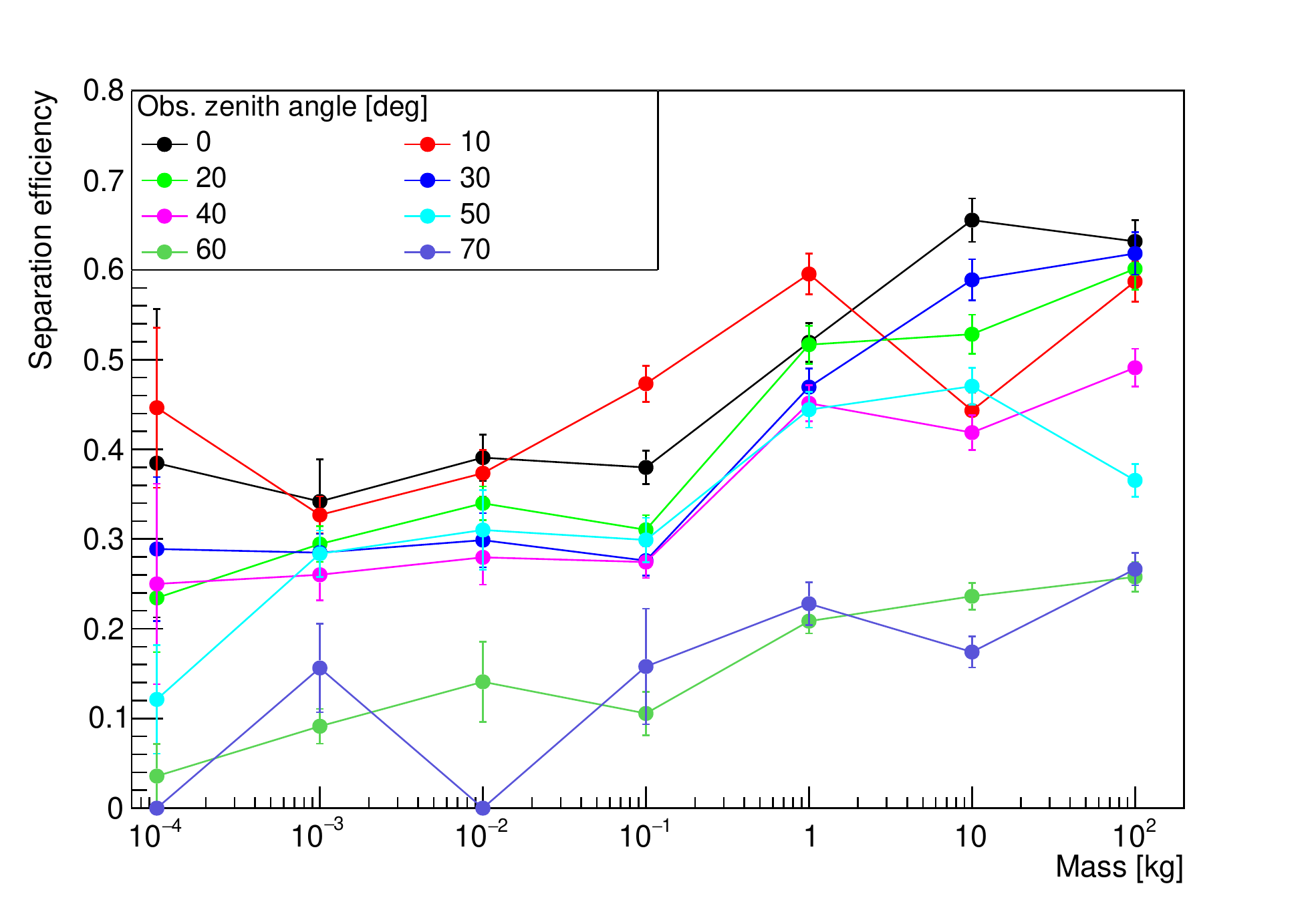}
		\caption{Separation efficiency - the fraction of simulated tracks remaining after removing all of the candidate events from the real data with a cut on length below 1060 pixels.}
		\label{fig:sep_efficiency}
	\end{center}
\end{figure}

If some of the false events have characteristics overlapping with those of heavy compact objects, a cut removing these fake events will also remove part of the objects' flux from our consideration. We call the fraction of the remaining tracks in all the simulated nuclearites a separation efficiency $\varepsilon_S$. Due to the long exposures, ``Pi of the Sky'' cannot base the separation on the measurement of the speed of the object. Also, the possible distinction based on the altitude of events would require stereoscopic observations. Therefore we have to rely on other characteristics of the recorded track.

Among several tested possibilities, a cut on the overall track length proved to be the most efficient one. In the case of nuclearites, the track should terminate (almost) only at the edges of the FoV, while for satellites and meteors, the exposure start and end, as well as light emission time should be the main limiting effects. The longest among the detected tracks remaining after the intensity profile variability based selection is about 1055 pixels long. Most tracks are shorter than 600 pixels (fig. \ref{fig:track_length}). If we assume that none of the detected tracks comes from a nuclearite, which seems reasonable as we have no other means to verify the origin of the events, the separation efficiency is above 50\% for heavy masses and pointing angles close to the zenith (fig. \ref{fig:sep_efficiency}) for which most tracks are bright. It naturally drops for dimmer tracks.

This a posteriori cut had to be used due to the lack of models representative for the whole population of possible background. Still, it gives similar results to the conservative ``maximum gap'' method \cite{bib:maximum_gap} of treating unknown background. A more detailed analysis of the distribution of the background events using the ``signal length''\cite{bib:signal_length} method shows that the probability of the events coming from the simulated track length distribution of nuclearites is lower than $10^{-6}$ (which cannot be, unfortunately, easily translated to an upper limit). Moreover, the same cut based on all the background events has been used for each nuclearite mass while, in theory, tracks could be distributed among nuclearite masses based on maximum brightness. Therefore our cut should be considered as an unorthodox, but simple way of estimating separation efficiency.

\subsection{Speed dependency}
\label{speed_dep}

The detection efficiency and thus the observable flux depends on the speed of nuclearite in two ways:

\begin{enumerate}
	\item The amount of light emitted by nuclearite in a unit of time is proportional to $v^3$
	\item The time spent in one pixel of the camera is proportional to $1/v$
\end{enumerate}

\begin{figure}
	\begin{center}
		\includegraphics[width=0.5\textwidth]{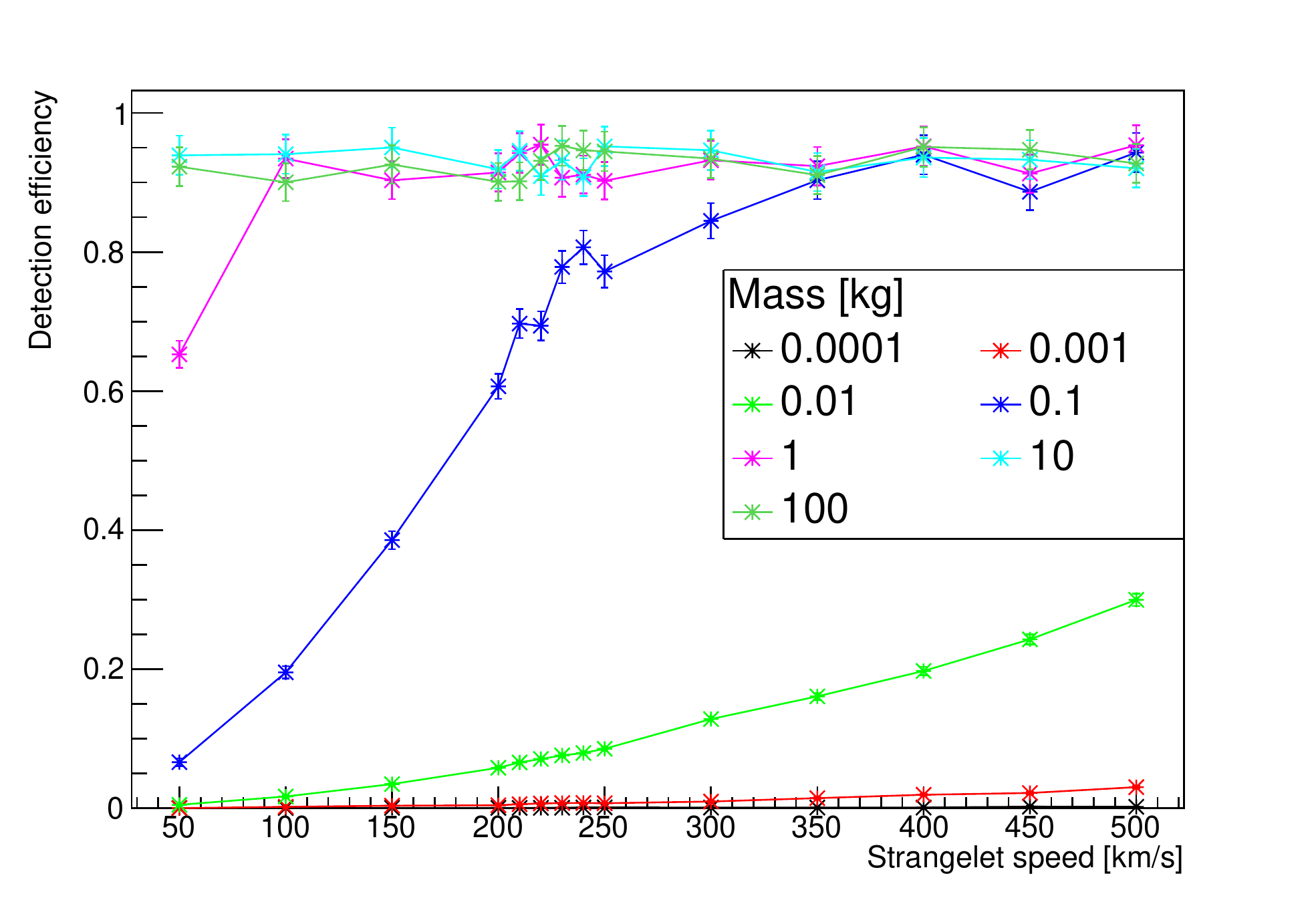}
		\caption{The detection efficiency vs the speed of the nuclearite}
		\label{fig:eff_vs_speed}
	\end{center}
\end{figure}

Therefore the brightness of a nuclearite in our data is proportional to $v^2$. For each considered mass, we have estimated the detection efficiency for speeds between 50 and 500 km/s in increments of 50 (fig. \ref{fig:eff_vs_speed}). Between 200 and 250 km, the step was 10 km/s, and linear interpolation was used for the directional flux calculation. The dependency shows that with increasing speed more and more masses will become visible to our detector. Of course, this is valid as long as the cross-section and the process of light emission remain the same.

\bibliography{z_methods}